# MERCADO E DESEMPENHO OPERACIONAL CONTÁBIL DE LONGO PRAZO

*MARKET AND LONG TERM OPERATIONAL ACCOUNTING PERFORMANCE*

**MEG SARKIS SIMÃO ROSA**
megsarkis@gmail.com

**PAULO ROBERTO BARBOSA LUSTOSA**
prblustosa@gmail.com

**RESUMO**

Seguindo as pesquisas sobre relevância da informação contábil para o valor da empresa (em inglês, categorizadas pela expressão *value relevance*), este estudo verifica se o mercado diferencia as empresas de alto, médio e baixo desempenho operacional de longo prazo, medido por informações contábeis de rentabilidade, variação de vendas e endividamento. Os dados compreendem as Demonstrações Contábeis Societárias Anuais Individuais divulgadas no período de 1996 a 2009 e o preço das ações das empresas listadas na Bolsa de Valores e de Mercadorias e Futuros de São Paulo – BM&FBOVESPA. A amostra final é composta por 142 empresas não financeiras. Foram utilizadas janelas móveis de cinco anos, que resultaram em dez períodos quinquenais. Após a apuração dos índices de cada empresa, as variáveis contábeis foram unificadas em um Índice Síntese de Desempenho para sintetizar a *performance* final por empresa a cada quinquênio, o que permitiu a segregação em níveis de desempenho operacional. Foram realizadas regressões múltiplas com técnicas de dados em painel, modelo de efeitos fixos e variáveis *dummies*, e depois realizados testes de hipóteses. Considerando o poder explicativo individual de cada variável, os resultados demonstraram que nem todos os comportamentos estão de acordo com as hipóteses da pesquisa e que o mercado acionário brasileiro diferencia empresas de alto e baixo desempenho operacional de longo prazo. Essa distinção não é percebida por completo entre as empresas de alto e médio desempenho operacional.

**Palavras-chave:** desempenho operacional contábil, retorno das ações, sustentabilidade de longo prazo.

**ABSTRACT**

Following the value relevance literature, this study verifies whether the marketplace differentiates companies of high, medium, and low long-term operational performance, measured by accounting information on profitability, sales variation and indebtedness. The data comprises the Corporate Financial Statements disclosed during the period from 1996 to 2009 and stock prices of companies listed on the São Paulo Stock Exchange and Commodities and Futures Exchange – BM&FBOVESPA. The final sample is composed of 142 non-financial companies. Five year mobile windows were used, which resulted in ten five-year periods. After checking each company's indices, the accounting variables were unified in an Index Performance Summary to synthesize the final performance for each five-year period, which allowed segregation in op-erational performance levels. Multiple regressions were performed using panel data techniques, fixed effects model and dummies variables, and then hypothesis tests were made. Regarding the explanatory power of each individual variable, the results show that not all behaviors are according to the research hypothesis and that the Brazilian stock market differentiates companies of high and low long-term operational performance. This distinction is not fully perceived between companies of high and medium operational performance.

**Key words:** accounting metrics of operational performance, stock returns, long-term sustainability.

## INTRODUÇÃO

O desempenho empresarial, que pode ser inferido com base em informações contábeis, revela a situação econômica das empresas. Avaliar esse desempenho é o objetivo dos investidores ao utilizarem a informação contábil para subsidiar decisões sobre compra, manutenção e venda de ativos na formação e administração dos portfólios de investimentos, consistentes com os riscos e retornos esperados. No entanto, como o mercado precifica os papéis em termos do seu valor econômico, a influência da informação contábil seria maior quanto mais esta se aproximasse do conceito de lucro econômico, definido como a diferença de valor entre dois momentos no tempo.

O lucro mensurado contabilmente é definido por um conjunto de regras apoiadas no conceito de realização financeira da receita (Hendriksen e Van Breda, 1999). No curto prazo, ele é bastante diferente do lucro mensurado em termos econômicos, porém, à medida que se expande a janela de tempo para mensuração do lucro contábil, mais este se aproxima da sua contraparte econômica. No limite, conforme observa Martins (1990), o lucro contábil é igual ao lucro econômico em toda a vida da empresa, a menos da inflação e do custo do capital próprio, pois essas variáveis não são incorporadas no lucro contábil.

A associação entre lucro e retorno das ações pode se apresentar baixa no curto prazo devido à presença de eventos econômicos que provocam revisões nas expectativas de mercado e que não são capturados nos resultados correntes, ou eventos econômicos passados podem estar influenciando os resultados atuais. Conforme Lev (1989), a arbitrariedade de várias medidas contábeis e técnicas de avaliação influenciam a relação lucro/retorno, porém, na média, o impacto dessas técnicas e manipulações de resultados diminui à medida que aumenta o período sobre o qual os lucros são mensurados.

O melhor desempenho sustentado de longo prazo poderia levar à suposição, pela hipótese de mercado eficiente, de que o mercado reconheceria maiores retornos acionários para as empresas de melhor desempenho. Porém, a mesma eficiência do mercado seria impeditiva que houvesse tal diferenciação, pois empresas mais rentáveis no mercado atrairiam a atenção de vários investidores dispostos a comprarsas ações dessas empresas, trazendo o retorno para a média do mercado.

A comprovação empírica das relações entre a informação contábil e a informação econômica de mercado, incorporada no preço dos papéis, é vasta na literatura acadêmica, sendo exemplos: Rowe e Morrow (1999); Kothari (2001); Santos e Lustosa (2008); Galdi e Lopes (2008); Almeida *et al.* (2008) e Malacrida (2009). Contudo, os estudos nacionais e estrangeiros dão maior ênfase à informação contábil de curto prazo, em janelas de tempos trimestrais, semestrais ou anuais.

Como os eventos econômicos correntes não são completamente refletidos no resultado de forma imediata, bem como eventos econômicos passados influenciam resultados correntes, o presente estudo considera que os reflexos dos eventos econômicos são percebidos nos resultados quando analisados no longo prazo e que eles contêm informações úteis ao mercado. Nesse sentido, parece relevante examinar o comportamento do mercado de capitais frente às informações contábeis em uma janela de tempo maior. Por essa razão, este trabalho se propõe a investigar a associação direta do preço das ações com o desempenho operacional de longo prazo, no sentido de responder a seguinte questão: *como o mercado diferencia as empresas de alto, médio e baixo desempenho operacional, medido por informações contábeis de retorno, variação de vendas e endividamento quando analisadas no longo prazo?*

Dada a questão anunciada, o objetivo geral da pesquisa é verificar como o mercado diferencia o desempenho operacional de longo prazo das empresas, medido por informações contábeis de diferentes naturezas, que sintetizam o desempenho quinquenal em empresas de alto, médio e baixo desempenho. Utilizando-se a técnica de dados em painel, este estudo é aplicado a certas variáveis contábeis extraídas de empresas com Demonstrações Contábeis Societárias Anuais Individuais divulgadas no período de 1996 a 2009, e retorno das respectivas ações listadas na Bolsa de Valores e de Mercadorias e Futuros de São Paulo – BM&FBOVESPA, conforme critério de seleção detalhado na metodologia. Assim, os resultados dessa pesquisa ficam restritos às empresas presentes na amostra. Os parâmetros de desempenho operacional são apurados a partir de dados contábeis anuais disponíveis no sítio da Comissão de Valores Mobiliários (CVM). Já os parâmetros de retorno das ações foram extraídos do banco de dados da Economática®, compreendendo o período de 1995 a 2009, sendo que o ano de 1995 foi tomado como ano-base para os cálculos das variáveis no primeiro quinquênio.

O presente artigo está estruturado da seguinte forma: além desta introdução, a segunda parte traz a fundamentação teórica que vem suportando os estudos sobre a relação existente entre variáveis contábeis e valores de mercado, bem como faz referências a algumas pesquisas relativas a esse tema, além de apresentar as hipóteses de pesquisa; a terceira seção demonstra os procedimentos de pesquisa; a quarta evidencia os resultados encontrados; e na última são apresentadas as considerações finais.

## FUNDAMENTAÇÃO TEÓRICA

A partir das contribuições de Ball e Brown (1968), a literatura contábil comprova a existência da relação entre a informação contábil e o preço das ações, utilizando-se do paradigma de que a contabilidade fornece informações úteis ao mercado de capitais. Os estudos sobre a relevância da informação contábil para o valor da empresa analisam o impacto do desempenho operacional, incorporado na informação contábil, no seu valor de mercado ou variáveis que capturam expectativas do mercado em relação ao desempenho e valor (Almeida *et al.*, 2008).

As pesquisas sobre relevância da informação contábil para o valor de mercado das empresas tipicamente incorporam argumentos contextuais contábeis para predizer a relação entre variáveis contábeis e valor de mercado (Beaver, 2002); operacionalizam o atributo de confiabilidade em termos da mensuração do erro; e procuram determinar a extensão das medidas de erros nos valores contábeis particulares, o que os torna operacionalizações empíricas dos critérios de relevância e confiabilidade afirmados pelo FASB (*Financial Accounting Standards Board*) no SFAC 5, o qual destaca que "informação contábil é relevante se for capaz de tornar diferentes as decisões dos usuários das demonstrações financeiras [...]; a informação contábil é confiável se representar o que propõe representar" (FASB, 1984, tradução nossa). Esses critérios também são ressaltados pelo IASB (*International Accounting Standards Board*), que considera que uma informação é relevante "quando influencia decisões econômicas dos usuários, ajudando-os a avaliar eventos passados, presentes ou futuros ou confirmando, ou corrigindo suas avaliações passadas" (IASB, 2010, p. A27).

Nota-se que os estudos sobre relevância para o valor levam em consideração um equilíbrio no mercado de capitais, tendo como foco os ativos e passivos individuais e os componentes de resultados, não o valor da firma com um todo. Portanto, a literatura sobre relevância para o valor não pode ser vista como única fonte de informação para usuários, acadêmicos ou não. Barth *et al.* (2001) ressaltam que não existe um modelo bem aceito de precificação de ativos em mercados imperfeitos e incompletos. Assim, pesquisas positivas que tratam da relação entre a contabilidade e o mercado de capitais utilizam conceitos de pesquisas empíricas em finanças, trazendo para a contabilidade o conceito da eficiência do mercado.

Fama (1970, p. 383) define um mercado eficiente como aquele em que os "preços das ações sempre 'refletem completamente' todas as informações disponíveis" e classifica essa eficiência em fraca, semiforte e forte, de acordo com os subconjuntos de informações disponíveis. Sob essa ótica, para que todas as informações relevantes sobre os eventos ocorridos possam influenciar no retorno dos investimentos, é necessário que as mesmas fiquem disponíveis aos seus participantes de maneira rápida e eficiente.

As pesquisas sobre a Hipótese da Eficiência do Mercado sugerem que os preços das ações não são influenciados apenas pelas informações contábeis, o que tem levado pesquisadores a investigar como as informações contábeis estão relacionadas aos retornos de mercado. A variação nos preços das ações, sob o contexto da Hipótese de Mercado Eficiente (HME), sinaliza a sensibilidade do mercado diante de novas informações, o que indica que a publicação das demonstrações contábeis, como informações relevantes, é útil na formação das suas expectativas. A base para as pesquisas que relacionam a informação contábil e o mercado de capitais consiste na premissa de que o mercado precifica as ações respaldado na informação disponível publicamente, isto é, considera que os mercados são eficientes na forma semiforte (Watts e Zimmerman, 1986).

O *Capital Asset Pricing Model* (CAPM), desenvolvido por Sharpe (1964), postula que o retorno de um ativo é igual à taxa de retorno livre de risco mais um prêmio de mercado pelo risco, medido pelo produto do preço de risco do mercado e o risco sistemático, o qual representa a volatilidade do retorno do ativo em resposta à volatilidade do retorno da carteira. A taxa esperada do retorno das ações é aumentada na covariância de risco do seu fluxo de caixa, isto é, a covariância de um retorno esperado da ação com o retorno esperado no portfólio do mercado. Assim, como no CAPM o retorno esperado de um ativo depende do seu risco sistemático, é possível estabelecer uma relação direta entre a informação contábil da empresa e seu risco (Watts e Zimmerman, 1986; Kothari, 2001).

A teoria que relaciona os relatórios contábeis ao valor de mercado da empresa depende de três suposições sobre as informações contidas nos resultados e nos preços das ações: (i) a contabilidade fornece informações para os participantes do mercado acerca da lucratividade atual e futura das empresas; (ii) essa lucratividade fornece, àqueles usuários, informações sobre os dividendos atuais e futuros esperados, ou seja, fluxos de caixa atuais e futuros esperados; (iii) o valor do preço das ações é igual ao valor presente dos dividendos futuros esperados, isto é, dos fluxos de caixa futuros esperados (Nichols e Wahlen, 2004). Nesse sentido, variações nos resultados sugerem alterações nas expectativas de fluxos de caixa dos participantes do mercado, o que implica variações nos preços das ações e, portanto, no valor de mercado da empresa.

Sendo assim, o modelo CAPM, do qual deriva a Hipótese de Eficiência do Mercado, permite a estimativa dos componentes de retorno da empresa. E, em conjunto com a Teoria da Divulgação, defendida por Verrechia (2001), o CAPM pode fornecer explicações e predições para as práticas contábeis e sua influência no valor da empresa no mercado acionário.

A Teoria da Divulgação tem como principal objetivo a explicação dos fenômenos relacionados à divulgação da informação financeira por meio de diversos aspectos como, por exemplo, determinar o efeito da divulgação das demonstrações financeiras nos preços das ações (Salotti e Yamamoto, 2005). O objetivo do trabalho de Verrechia (2001) é caracterizar os vários modelos que tratam da divulgação, no qual o autor propõe uma taxonomia em que destaca três categorias de pesquisas sobre divulgação em contabilidade: (i) pesquisa sobre divulgação baseada em associação (*association-based disclosure*); (ii) pesquisa sobre divulgação baseada em julgamento (*discretionary-based disclosure*); (iii) pesquisa sobre divulgação baseada em eficiência (*efficiency-based disclosure*).

Salotti e Yamamoto (2005) corroboram Verrechia (2001) ao afirmarem que as pesquisas sobre divulgação, baseadas na associação, buscam examinar a relação entre o fenômeno da divulgação e as mudanças no comportamento dos investidores, para maximizar suas riquezas individuais. Como o objetivo desta pesquisa é verificar se há associação direta do retorno das ações com o desempenho contábil-financeiro das empresas, pode-se enquadrá-la na categoria de pesquisas baseadas em associação, na classificação desenvolvida por Verrechia (2001). Assim, de acordo com a Teoria da Divulgação, espera-se que informação contábil seja relevante à medida que possa influenciar os investidores e o próprio mercado de capitais.

### *HIPÓTESES DA PESQUISA*

Ao pesquisar *performance* passada, mensurada por meio de dados contábeis, como previsora para *performance* futura baseada em dados de mercado, Hoskisson *et al.* (1994 *in* Rowe e Morrow, 1999) encontraram um relacionamento significantemente positivo entre medidas de desempenho financeiro baseadas em indicadores contábeis e de mercado. Nesse contexto, Lyra e Corrar (2009) desenvolveram uma pesquisa junto a docentes brasileiros e norte-americanos sobre quais indicadores são necessários e suficientes para diferenciação e avaliação de desempenho. Dentre os indicadores mais citados estão: retorno sobre patrimônio líquido, classificado por unanimidade como o principal indicador de desempenho empresarial; rentabilidade sobre o ativo; crescimento das vendas; liquidez corrente; composição do endividamento; margem líquida; e giro do ativo.

A importância dos lucros para o mercado é reconhecida por autores como Dechow *et al.* (1998) ao afirmarem que lucros ocupam uma posição central na contabilidade por refletir o fluxo de caixa projetado e ter uma maior relação com o fluxo de caixa corrente. E, segundo Hendriksen e Van Breda (1999), conhecer os fluxos de caixa futuros esperados é o que permite o mercado fixar os preços das ações de uma empresa.

Os estudos de Santos e Lustosa (2008) verificaram qual a métrica do resultado contábil trimestral que melhor expressa os fatores considerados pelos participantes do mercado na formação do preço das ações e os resultados indicaram que o lucro líquido é a métrica que melhor capta e expressa os valores considerados na formação dos preços das ações. Porém, Lev (1989) avaliou a utilidade dos lucros para os investidores e encontrou uma correlação muito baixa entre lucro e retorno das ações. O autor afirma que a natureza dessa relação exibe consideráveis instabilidades ao longo do tempo, existindo uma diferença significativa na relação lucro/retorno no curto prazo em relação ao longo prazo; quando se expande a janela temporal dos estudos, os coeficientes de correlação ($R^2$) tendem a aumentar.

A relação contemporânea entre lucro e retorno das ações cresce fortemente se analisada no longo prazo devido à presença de eventos econômicos que provocam revisões nas expectativas de mercado e que não são capturados nos resultados correntes, ou eventos econômicos passados podem estar influenciando os resultados atuais. Com o objetivo de investigar a relevância da informação contábil para o mercado de capitais de países emergentes, Galdi e Lopes (2008) encontraram evidências de um relacionamento de longo prazo entre o lucro contábil e o preço das ações para a maior parte das empresas analisadas.

Os fluxos de caixa que entram e saem das empresas também são eventos fundamentais nos quais as mensurações contábeis são baseadas e que se supõem que os investidores utilizam para respaldar suas decisões (Hendriksen e Van Breda, 1999). Assim, o retorno sobre o investimento em termos de fluxo de caixa (RIFC) representa o fluxo de caixa sustentável que a empresa gera em suas operações em determinado período como um percentual do caixa investido. Malacrida (2009) analisou a relevância do fluxo de caixa operacional corrente, *accruals* e lucros correntes para predizerem os fluxos de caixa operacionais futuros e estimarem os retornos das ações na Bolsa de Valores de São Paulo; os resultados demonstraram que o fluxo de caixa operacional corrente possui maior capacidade preditiva do fluxo de caixa futuro do que os lucros correntes.

Frigo (2008) ressalta que uma empresa é considerada de elevado desempenho quando apresentar habilidade para alcançar um retorno sobre o investimento elevado e mantê-lo sustentável por um período de 10 anos consecutivos ou mais; deve, também, aumentar os negócios enquanto mantém o retorno sustentável e, por consequência, a combinação dessas duas dimensões leva a um retorno superior para o acionista.

A sustentabilidade pode ser percebida, em termos práticos, pelo desempenho operacional da empresa em janelas de tempo mais longas. Nesse sentido, anuncia-se a seguinte hipótese, na forma alternativa:

$H_1$: O mercado acionário brasileiro reage positivamente às informações contábeis que denotam rentabilidade e crescimento, e esta reação ocorre de modo diferenciado conforme o nível de desempenho operacional de longo prazo das empresas.

O modelo de *Pecking Order* proposto por Myers (1984) defende que a assimetria da informação tem influência nas decisões de financiamentos, o que privilegia a posição dos gerentes ao buscarem financiamentos com ativos arriscados quando a empresa está valorizada. Porém, a racionalidade dos financiadores faz com que estes descontem o valor da firma quando ativos arriscados são utilizados para novos investimentos. Essa situação desagrada o mercado, que interpreta uma estrutura de capital elevada como uma má notícia, influenciando nos preços das ações.

A Teoria de *Pecking Order* estabelece, ainda, uma hierarquia de escolhas que as empresas devem seguir sobre o financiamento de seus projetos. Recorrendo, primeiramente, a recursos gerados internamente, seguidos de emissão de dívidas e por último, caso necessário, emissão de novas ações. Assim, as empresas procuram financiar seus investimentos com recursos próprios, na expectativa de que o endividamento decresça quando investimentos não excedem os lucros. Portanto, existe uma relação positiva entre investimento e dívida e uma relação negativa entre dívida e lucro, ou seja, quanto menos endividada, maior será a lucratividade da empresa e maiores serão seus dividendos.

Existem fatores que inibem ou estimulam o endividamento das empresas. Alguns estão relacionados à economia em geral e outros estão relacionados aos aspectos internos, como características operacionais e desempenho. Titman e Wessels (1988) utilizaram análise fatorial e observaram que empresas com baixo desempenho tendem a acumular dívidas para suportar suas despesas. Este estudo foi replicado no Brasil por Famá e Perobelli (2001), que encontraram relação negativa entre crescimentos dos ativos, lucratividade e tamanho da empresa com o grau de endividamento de curto prazo.

Silva e Brito (2003) observaram que as empresas brasileiras mais lucrativas e as que investem menos são menos endividadas, conforme defende a Teoria. Assim, seguindo a Teoria da *Pecking Order*, apresenta-se a seguinte hipótese:

$H_2$: O mercado acionário brasileiro reage negativamente às informações contábeis que denotam endividamento, e esta reação é menor quanto maior for o desempenho operacional de longo prazo das empresas.

**PROCEDIMENTOS DA PESQUISA**

Os estudos foram realizados sobre janelas móveis de cinco anos, o que resultou em dez períodos quinquenais e uma amostra composta por 142 empresas com ações negociadas na Bolsa de Valores e de Mercadorias e Futuros de São Paulo (BM&FBOVESPA) que atenderam, cumulativamente, aos seguintes critérios de seleção: Média anual mínima de 120 negociações para cada quinquênio analisado;[1] Não pertencer ao grupo de empresas financeiras, seguradoras e fundos; Não ser classificada como *Holding*; Apresentar ciclos operacionais de no máximo um ano;[2] e Disponibilizar dados completos no quinquênio, sendo que o primeiro quinquênio com dados completos foi considerado como base de cálculo para o quinquênio subsequente.

Os dados contábeis foram obtidos em março de 2010, por meio do sítio da Comissão de Valores Mobiliários (CVM) e corrigidos monetariamente pelo Índice Nacional de Preços ao Consumidor Amplo (IPCA) para fevereiro de 2010, abrangendo observações individuais e anuais entre 1995 e 2009. No que compete às ações, foi considerado o valor de fechamento no último dia de cotação do ano, ajustados pelos dividendos. Os preços foram extraídos da base de dados Economática®. O critério de escolha do tipo de ação a ser estudado foi o de maior liquidez no período que compreende a pesquisa.

***MODELO ECONOMÉTRICO***

A aplicação dos testes empíricos considera a especificação de uma regressão linear múltipla, a partir da qual seja possível estimar a reação do mercado diante do desempenho operacional das empresas no longo prazo. Utilizou-se uma estrutura de dados em painel com efeitos fixos, determinado pelo *Teste de Hausman*, que verifica a existência de correlação entre os efeitos individuais e as variáveis explicativas, conforme ressalta Wooldrigde (2006).

As variáveis contábeis escolhidas para representar o desempenho operacional mediante índices de rentabilidade, crescimento da empresa e o seu endividamento foram: Retorno sobre Patrimônio Líquido (RSPL), Retorno sobre Investimento em termos de Fluxo de Caixa (RIFC), Crescimento das Receitas (REC) e a Estrutura de Capital (EC), as quais foram calculadas tomando por base períodos de cinco anos, considerando as metodologias específicas de cada uma.

O Retorno sobre o Patrimônio Líquido (RSPL) representa a taxa de rentabilidade auferida pelo capital próprio da empresa, o que propicia mensurar, entre outras coisas, se a alavancagem financeira está produzindo ou destruindo valor das empresas. Este índice foi mensurado pelo quociente entre o lucro líquido quinquenal (depois de juros e impostos) e a média aritmética simples entre os patrimônios líquidos inicial e final do quinquênio.

---

[1] O investidor cobra prêmio de iliquidez para adquirir ações menos líquidas. Consequentemente, na média, ações mais líquidas têm maior valor de mercado. Machado e De Medeiros (2011), utilizando cinco diferentes medidas de liquidez de ações, confirmaram a existência do prêmio de iliquidez em ações negociadas na BM&FBOVESPA.

[2] A limitação do tamanho do ciclo operacional em um ano implica maior quantidade de ciclos no período quinquenal, caracterizando mais adequadamente a ideia de sustentabilidade do desempenho contábil e sua relação com o mercado acionário investigada nesta pesquisa.

Já o retorno sobre o investimento em termos de fluxo de caixa (RIFC) foi calculado pelo quociente entre o Fluxo de caixa operacional do quinquênio e o ativo médio operacional de cada quinquênio, revelando o quanto a empresa está ganhando em suas operações (em termos de caixa) em relação aos ativos investidos. Cabe ressaltar, no entanto, que o fluxo de caixa operacional calculado é uma *proxy* aproximada do seu valor verdadeiro sendo deixadas de fora apenas algumas contas do ativo e passivo circulante que não pertencem ao fluxo de caixa das operações, sendo exemplos os dividendos a pagar e a parcela de curto prazo de empréstimos bancários e de emissão de dívidas (atividades de financiamento).

O crescimento dos negócios, nesta pesquisa, é representado pela variável Crescimento da Receita Líquida (REC). Para predizer o efeito da estrutura de capital, o estudo adota como premissa a teoria *Pecking Order*, por ser considerada como uma *proxy* de risco de uso de capitais de terceiros. A estrutura de capital foi mensurada pela razão entre o endividamento total no fim de cada quinquênio e o ativo total médio do quinquênio.

Visando controlar o efeito do mercado sobre o retorno das ações, foi utilizado o Retorno do Índice Bovespa (IBOV) como variável de controle. Para representar o comportamento do mercado nos períodos quinquenais, o IBOV foi obtido através de capitalização contínua, pelo logaritmo natural da razão entre o IBOV do início e fim de cada quinquênio. Como o objetivo do estudo é verificar se há associação significativa e direta do retorno das ações no mercado com o desempenho operacional das empresas, medido por métricas contábeis e considerando períodos de tempo quinquenais, o método de capitalização contínua também foi utilizado para o cálculo do retorno das ações.

A fim de apresentar coeficientes de respostas específicos de cada grupo de empresas (alto, médio e baixo desempenho), aplicou-se o modelo de Efeitos Fixos com variáveis *dummies*. Nesse sentido, construiu-se o seguinte modelo econométrico:

$$R_{it} = \beta_o + \beta_{1j}RSPL_{it} + \beta_{2j}d2RSPL_{it} + \beta_{3j}d3RSPL_{it} + \beta_{4j}RIFC_{it} + \beta_{5j}d2RIFC_{it} + \beta_{6j}d3RIFC_{it} + \beta_{7j}REC_{it} + \beta_{8j}d2REC_{it} + \beta_{9j}d3REC_{it} + \beta_{10j}EC_{it} + \beta_{11j}d2EC_{it} + \beta_{12j}d3EC_{it} + \beta_{13j}IBOV_t + u_{it}$$

Onde,
$R_{it}$ = Retorno da ação da empresa i, no quinquênio t;
$\beta_o$ = Intercepto geral;
$\beta_{1j}$ = Coeficiente de resposta da variável RSPL, do grupo (tercil) j;
$RSPL_{it}$ = Retorno sobre Patrimônio Líquido da empresa i, no quinquênio t;
$\beta_{2j}$ e $\beta_{3j}$ = Coeficientes de resposta das variáveis com *dummies* multiplicativas d2RSPL e d3RSPL, respectivamente, do grupo (tercil) j;
$d2[.]_{it}$ e $d3[.]_{it}$ = *Dummies* multiplicativas para agrupar à subamostra do primeiro tercil os tercis 2 e 3, respectivamente, para cada uma das 4 variáveis contábeis (RSPL, RICF, REC e EC) tratadas na pesquisa, da empresa i, no quinquênio t;
$\beta_{4j}$ = Coeficiente de resposta da variável RIFC, do grupo (tercil) j;
$RIFC_{it}$ = Retornos sobre Investimento em termos de Fluxo de Caixa da empresa i, no quinquênio t;
$B_{5j}$ e $\beta_{6j}$ = Coeficientes de resposta das variáveis com *dummies* multiplicativas d2RIFC e d3RIFC, respectivamente, do grupo (tercil) j;
$\beta_{7j}$ = Coeficiente de resposta da variável REC do grupo (tercil) j;
$REC_{it}$ = Crescimento da Receita da empresa i, no quinquênio t;
$\beta_{8j}$ e $\beta_{9j}$ = Coeficientes de resposta das variáveis com *dummies* multiplicativas d2REC e d3REC, respectivamente, do grupo (tercil) j;
$\beta_{10j}$ = Coeficiente de resposta da variável EC do grupo (tercil) j;
$EC_{it}$ = Estrutura de Capital da empresa i, no fim do quinquênio t;
$\beta_{11j}$ e $\beta_{12j}$ = Coeficientes de resposta das variáveis com *dummies* multiplicativas d2EC e d3EC, respectivamente, do grupo (tercil) j;
$\beta_{13j}$ = Coeficiente de resposta da variável de controle IBOV do grupo (tercil) j;
IBOV = Retorno do índice Bovespa, no quinquênio t;
$u_{ijt}$ = termo de erro da regressão.

## SEGREGAÇÃO DA AMOSTRA

Após apurados os índices de cada empresa, as variáveis independentes (exceto a de controle) foram unificadas em um único Índice Síntese de Desempenho (ISD), por meio de atribuições de pesos a cada variável contábil de desempenho operacional, como *proxy* para sintetizar a *performance* final por empresa a cada quinquênio:

$$ISD_{it} = 0{,}5 * RSPL_{it} + 0{,}2 * RICF_{it} + 0{,}2 * REC_{it} + 0{,}1 * EC_{it}$$

Onde,
$ISD_{it}$ = Índice síntese de desempenho da empresa i, no quinquênio t.
$RSPL_{it}$ = Retorno sobre Patrimônio Líquido da empresa i, no quinquênio t.
$RICF_{it}$ = Retorno sobre o Investimentos em Termos de Fluxo de Caixa da empresa i, no quinquênio t.
$REC_{it}$ = Crescimento da Receita da empresa i, no quinquênio t.
$EC_{it}$ = Estrutura de capital da empresa i, no quinquênio t.

Estes pesos foram determinados com base na importância dada a essas variáveis em estudos anteriores, como por exemplo, nos estudos que tratam o lucro como uma posição central na contabilidade (Dechow *et al.*, 1998; Galdi e Lopes, 2008; Santos e Lustosa, 2008; Lyra e Corrar, 2009).

Para segregar a amostra em tercis, visando classificar as empresas em alto, médio e baixo desempenho operacional no longo prazo, foi calculado o Índice Síntese Final de Desempenho (ISFD), que é representado pela média aritmética de todos os Índices Sínteses de Desempenhos (ISD) quinquenais da empresa.

Depois de calculado o ISFD de cada empresa, a amostra total foi classificada em ordem decrescente e segregada em tercis, sendo que, no primeiro tercil, situam-se as empresas com maior desempenho sustentável no longo prazo; no segundo, as de médio desempenho; e, no terceiro, as empresas que apresentaram baixo desempenho operacional no longo prazo. Este procedimento captura o desempenho móvel quinquenal de cada empresa. Ao apurar-se o ISFD pela média aritmética dos ISD nos 10 quinquênios, obtém-se um panorama das empresas que tiveram desempenhos melhores, intermediários e piores no conjunto dos 10 quinquênios, sendo este índice o indicador para gradar a amostra resultante pelos três grupos de desempenho operacional contábil de longo prazo.

***PROCEDIMENTO DE TESTE DE HIPÓTESES***

Tendo em vista a análise da associação entre o retorno das ações e as variáveis contábeis que denotam o desempenho operacional das empresas no longo prazo, os testes realizados, que têm como base o modelo econométrico antes definido, avaliaram se os coeficientes de respostas das variáveis contábeis das empresas são estatisticamente significantes em sua associação com o retorno das respectivas ações. Para verificar se os coeficientes de respostas do grupo (tercil) de maior desempenho operacional são significativamente diferentes em relação aos coeficientes dos grupos de empresas que apresentam médio e baixo desempenho, foi utilizado o Teste de *Wald*.

Os testes foram realizados, conforme modelo econométrico, por acréscimos sucessivos à subamostra do 1º tercil (empresas de melhor desempenho médio quinquenal) das subamostras do 2º tercil (desempenho médio quinquenal intermediário) e 3º tercil (pior desempenho médio quinquenal). Este procedimento pode ser especificado da seguinte forma genérica:

$$y = \beta 1x + \beta 2xd_2 + \beta 3xd_3$$

onde *y* representa o retorno das ações e *x* cada variável contábil, sendo d1 e d2 *dummies* multiplicativas para as subamostras do 2º e 3º tercil, respectivamente.

Para a realização dos testes com as empresas do 1º tercil, d1 e d2 são 0 (zero), e o coeficiente de resposta das variáveis contábeis é β1. Em seguida, acrescenta-se a subamostra das empresas do 2º tercil à subamostra do 1º tercil, fazendo-se d2 = 1 e d3 = 0. Com isso, os novos coeficientes de resposta das variáveis contábeis serão a soma algébrica de β1 e β2, pois a especificação geral será:

$$y = \beta 1x + \beta 2(x*1) + \beta 3(x*0)$$
$$\text{ou } y = (\beta 1 + \beta 2)x$$

Analogamente, ao se comparar o 1º tercil com o 3º tercil, faz-se d2 = 0 e d3 = 1. Com isso, os novos coeficientes de resposta das variáveis contábeis serão a soma algébrica de β1e β3, pois a especificação geral será:

$$y = \beta 1x + \beta 2(x*0) + \beta 3(x*1)$$
$$\text{ou } y = (\beta 1 + \beta 3)x$$

Para que os novos coeficientes de resposta das variáveis RSPL, RICF e REC, em sua associação com os retornos e após a agregação do 2º e 3º tercis, sejam menores do que os do 1º tercil, espera-se um efeito negativo desses tercis de menor desempenho nos retornos, já que, por exemplo, o beta da 2ª subamostra será a soma algébrica do beta do 1º tercil (β1) com o efeito produzido pela agregação do 2º tercil à 1ª amostra, β22 = β11 + β12. Assim, β22 < β11 ↔ β12 < β11. Um efeito contrário é esperado para o coeficiente de resposta da variável EC, pois espera-se que o acréscimo alternado das subamostras de menor desempenho (2º e 3º tercis) ao 1º tercil potencialize a associação negativa entre os retornos e a estrutura de capital.

Nesse sentido, são testadas as seguintes hipóteses:

$H_1$:

$$\beta_{1j}(j = 1) > \beta_{2j}(j = 2) > \beta_{3j}(j = 3) > 0$$
$$\beta_{4j}(j = 1) > \beta_{5j}(j = 2) > \beta_{6j}(j = 3) > 0$$
$$\beta_{7j}(j = 1) > \beta_{8j}(j = 2) > \beta_{9j}(j = 3) > 0$$

$H_2$:

$$\beta_{10j}(j = 1) < \beta_{11j}(j = 2) < \beta_{12j}(j = 3) < 0$$

***TESTES DE ROBUSTEZ***

Para medir a robustez dos dados empíricos, foram realizados testes quanto à existência de raiz unitária nas séries, quanto à distribuição das séries e quanto à existência de autocorrelação nos termos de perturbação aleatórios. A heterocedasticidade em dados de corte transversal pode ser mais a regra do que a exceção (Gujarati, 2006). Assim, devido ao programa estatístico utilizado (Eviews® 6.0) não realizar teste de heterocedasticidade em dados em painel, optou-se por utilizar erros-padrão com a correção da heterocedasticidade de *White*, também conhecidos como erros-padrão robustos. As variáveis RSPL e RICF apresentaram raízes unitárias. Esse problema foi corrigido com o uso da 1ª diferença destas variáveis (ΔRSPL e ΔRICF), em lugar das medidas em nível, nas regressões.

**RESULTADOS**

A amostra foi composta por 142 empresas selecionadas para estudo, distribuídas, concomitantemente ou não, em dez períodos quinquenais que atenderam aos critérios de seleção adotados. A média de empresas por quinquênio foi de, aproximadamente, 80 empresas, totalizando 695 observações. As empresas foram classificadas em ordem decrescente de desempenho calculado pelo ISFD, o maior desempenho sustentável no período em estudo foi de 2,7108, enquanto que o menor desempenho do período foi -10,1604.[3]

Com a estatística descritiva calculada ao nível, isto é, antes dos testes de raiz unitária, a distribuição apresentou-se levemente positiva para as empresas de alto e médio desempenho, bem como para os dados dispostos em conjunto. Porém, quando analisadas as empresas de baixo desempenho, percebeu-se uma inclinação negativa para todas as variáveis, com exceção da estrutura de capital que permaneceu com a inclinação levemente positiva.

A média quinquenal do retorno das ações para as empresas de alto desempenho foi de 75,34%, enquanto que para as empresas de médio desempenho foi de 49,59%, o equivalente a uma redução de 34,18%. Essa diferença aumentou quando comparados alto e baixo desempenhos (-7,95% de retorno médio no quinquênio), diminuindo aproximadamente 110,55%, o que pode demonstrar preferência do mercado pelas empresas que apresentaram desempenho operacional sustentável no longo prazo.

O retorno sobre patrimônio líquido possui desvio-padrão elevado para as empresas de alto (99,78%) e baixo desempenho (135,39%), indicando, neste último, a presença de um pequeno grupo de empresas com forte tendência à destruição de valor para o acionista. Já o crescimento da receita e a estrutura de capital apresentaram menores variabilidades dos desvios e médias positivas nos três níveis de desempenho.

Para verificar o grau de associação entre as variáveis, foram calculados os coeficientes de correlação de *Pearson*, mostrados na Tabela 1. Os valores das correlações servem como uma referência preliminar das relações existentes entre as variáveis adotadas na pesquisa.

Na Tabela 1, encontram-se as correlações de *Pearson* entre as variáveis utilizadas no modelo testado. Pelos valores apresentados, todas as variáveis apresentam correlações significantes, para o nível de 1%, com a variável dependente (R).

A estrutura de capital apresentou correlação negativa com as demais variáveis, com exceção do crescimento das receitas, que se apresentou levemente correlacionado. Segue, portanto, a teoria de *Pecking Order* e corrobora o estudo de Famá e Parobelli (2001), no qual encontraram relação negativa entre a lucratividade e o grau de endividamento das empresas brasileiras.

Entre as variáveis independentes deve-se destacar a forte correlação entre o RSPL e o RIFC. Esta correlação era esperada, pois as magnitudes do fluxo de caixa operacional e do lucro líquido (este último se incorpora ao Patrimônio Líquido) tendem a se aproximar apesar da defasagem temporal entre os lucros apurados pelos regimes de competência e de caixa, já que as receitas e despesas que passaram pelo lucro em períodos anteriores são recebidas e pagas no período corrente, compensando as receitas e despesas do período corrente que somente serão recebidas e pagas no futuro (Lustosa e Santos, 2006).

Cabe ressaltar, no entanto, que a alta correlação esperada entre o retorno sobre o patrimônio líquido (RSPL) e retorno sobre investimento em termos de fluxo de caixa (RIFC) pode produzir algum efeito não desejado na associação entre a variável dependente e as variáveis independentes.

A correlação entre o crescimento da receita e o Ibovespa apresentou-se significante em 5%, porém, com sinal negativo, o que leva a sugerir que o crescimento da receita medido pelo mercado seja baseado em informações econômicas, como fluxos de caixa futuros, diferenciadas daquelas respaldadas no princípio da competência.

### ASSOCIAÇÃO DO RETORNO DAS AÇÕES COM AS VARIÁVEIS CONTÁBEIS

Os resultados demonstram que as variáveis contábeis que denotam desempenho operacional são significantemente associadas ao retorno das ações, ($R^2$ ajustado de 0,6499). Quan-

Tabela 1 - *Correlação de Pearson entre as variáveis.*
Table 1 – *Pearson correlation between variables.*

|      | R        | RSPL     | RIFC     | REC     | EC       | IBOV |
|------|----------|----------|----------|---------|----------|------|
| R    | 1        |          |          |         |          |      |
| RSPL | 0,399**  | 1        |          |         |          |      |
| RIFC | 0,212**  | 0,605**  | 1        |         |          |      |
| REC  | 0,204**  | 0,123**  | 0,110**  | 1       |          |      |
| EC   | -0,086** | -0,049** | -0,217** | 0,022   | 1        |      |
| IBOV | 0,328**  | 0,050    | 0,028    | -0,077* | -0,087** | 1    |

Nota: Os asteriscos indicam o nível de significância dos coeficientes *(5%) e **(1%).

---

[3] Por limitação de espaço, optou-se por apenas comentar narrativamente algumas estatísticas descritivas desta pesquisa, sem a apresentação das tabelas, uma vez que estas são bastante extensas. O leitor interessado pode obter os dados que desejar em contato com os autores.

do analisado o poder explicativo individual de cada variável, percebe-se que há, em termos gerais, um alinhamento com as hipóteses da pesquisa.

A Tabela 2 apresenta evidências de que o mercado realmente diferencia as empresas conforme o seu desempenho operacional de longo prazo (no caso, quinquenal), medido por métricas contábeis, mesmo depois que se controla pelo grande efeito (estatística t = 14,5617) que o retorno da Bovespa (IBOV) produz nos retornos do mercado acionário.

Analisando-se pela ótica do Retorno sobre o Patrimônio Líquido (RSPL), a medida contábil mais completa para sintetizar o desempenho operacional da empresa, o coeficiente de resposta dessa variável diminuiu consistentemente, como esperado, entre os três grupos de desempenho. No grupo de empresas de alto desempenho, o coeficiente de RSPL (trabalhou-se com ∆RSPL para corrigir o problema de raiz unitária com essa variável) foi de 0,4370, estatisticamente significativo a 99,99% de grau de confiança. Quando as empresas de desempenho intermediário foram agregadas às de alto desempenho, esperava-se uma redução na resposta do mercado, o que de fato aconteceu, pois o novo coeficiente de RSPL caiu quase 100%, para 0,0777, e essa variável passou a não mais ser significativa (estatística t = 0,4266) para explicar o retorno; e quando as empresas do grupo intermediário de desempenho foram substituídas pelas de pior desempenho e agregadas à subamostra de melhor desempenho (1º tercil), o coeficiente de RSPL caiu mais ainda, para um valor negativo de -0,3059, que é significativo a 99% de grau de confiança (estatística t = -2,7352). Isto significa que uma estratégia de investimento centrada em empresas com maior RSPL sustentado, que nesta pesquisa foi considerado como períodos móveis quinquenais ao longo de 10 anos, poderia gerar retornos anormais positivos aos investidores.

A análise pela ótica da variável RICF é menos contundente do que a do RSPL, pois o RICF é uma medida de retorno baseada no fluxo de caixa das operações, que é mais volátil do que o lucro. Por essa razão, a oscilação dos coeficientes de resposta do RICF entre os grupos de desempenho é maior, embora ainda tenha prevalecido a lógica de o mercado reconhecer as empresas de alto desempenho quinquenal como mais atrativas que as dos demais grupos. Para o grupo de alto desempenho, o coeficiente de RICF (considerou-se a variação dessa variável nas regressões, para corrigir problema de raiz unitária) foi de 0,5932, significativo a 99% (stat t = 2,7036). Significa que empresas com maior RICF terão maiores retornos no mercado acionário. Com a agregação da subamostra de empresas de desempenho intermediário à subamostra de empresas de alto desempenho, houve uma inversão do sinal do coeficiente de resposta de RICF, que passou a ser de -0,7725, significativo a 99%. Isto indica, como esperado, que, na média, quanto maior o *retorno negativo* do fluxo de caixa das operações em relação aos ativos da empresa (na verdade, um prejuízo), maior será o retorno negativo da ação da empresa no mercado acionário. Quando houve a agregação de empresas do grupo de pior desempenho (3º tercil) à subamostra do 1º tercil (alto desempenho), em substituição às empresas do 2º tercil, o coeficiente de RICF manteve-se negativo (-0,3572), mas não foi significativo (stat t = -0,8646). Em síntese, semelhantemente ao RSPL, uma estratégia de investimentos centrada em empresas com maior razão quinquenal entre o fluxo de caixa das operações e o ativo total operacional geraria maior retorno anormal positivo no mercado acionário para os investidores.

Tabela 2 - *Resumo do Resultado da Regressão.*
Table 2 – *Summary of Regression Results.*

|  | Variáveis | Coeficientes | Estatística t | p-valor |
|---|---|---|---|---|
| Alto Desempenho | ∆RSPL | 0,4370 | 4,7726 | 0,0010 |
|  | ∆RIFC | 0,5932 | 2,7036 | 0,0071 |
|  | REC | 0,7036 | 1,7434 | 0,0820 |
|  | EC | -0,0305 | -1,1627 | 0,2456 |
| Médio Desempenho | ∆RSPL | 0,0777 | 0,4266 | 0,6699 |
|  | ∆RIFC | -0,7725 | -3,1277 | 0,0019 |
|  | REC | 1,0803 | 1,6175 | 0,1065 |
|  | EC | -0,1009 | -0,9390 | 0,3483 |
| Baixo desempenho | ∆RSPL | -0,3059 | -2,7352 | 0,0065 |
|  | ∆RIFC | -0,3572 | -0,8646 | 0,3877 |
|  | REC | -0,6020 | -1,2293 | 0,2196 |
|  | EC | -0,0309 | -0,6969 | 0,4862 |
|  | IBOV | 0,7879 | 14,5617 | 0,0000 |
| R² Ajustado | 0,6499 |  |  |  |

Os resultados para a variável REC, que captura a variação quinquenal do faturamento das empresas da amostra, corroboraram parcialmente a lógica teórica desta pesquisa. O grupo de empresas do 1º tercil, com maior desempenho quinquenal medido por variáveis contábeis, apresenta coeficiente de resposta 0,7036, significativo a 90% de grau de confiança. Ao contrário do esperado, entretanto, a agregação das empresas do 2º tercil à primeira subamostra elevou o coeficiente de resposta da variável REC para 1,0803, significativo no limite do grau de confiança a 90%. Mas na junção das empresas do 1º tercil (alto desempenho) com as do 3º tercil (pior desempenho), o coeficiente de resposta de REC caiu significativamente, como esperado, para o valor negativo de -0,6020, embora este não tenha se revelado estatisticamente diferente de zero. Novamente, há alguma evidência de que estratégias de investimento em empresas com alto desempenho contábil de longo prazo, mesmo quando esse desempenho for medido pela variação quinquenal da receita, uma medida de desempenho bem mais restrita do que a variável RSPL, podem gerar retornos anormais positivos para os investidores.

Finalmente e também como esperado, o coeficiente de resposta da variável EC estrutura de capital (EC) apresentou-se negativo para os três grupos, -0,0305, -0,1009 e -0,0309, respectivamente. Houve um aumento da magnitude negativa (o que representa uma piora, como esperado) com a agregação da subamostra do 2º tercil ao 1º tercil, mas, contrariamente ao esperado, voltou aos níveis da 1ª subamostra quando esta foi juntada à 3ª subamostra. Mas, com respeito a essa variável (EC), os coeficientes de resposta dos três grupos de empresa, embora negativos, não são estatisticamente diferentes de zero.

### *RESULTADOS DOS TESTES DE HIPÓTESES*

Foram realizados testes de *Wald* para testar a hipótese de que o mercado de capitais trata de modo diferenciado as empresas de alto desempenho operacional sustentável no longo prazo. Os coeficientes de respostas (β) das variáveis contábeis de alto desempenho foram comparados com os respectivos coeficientes das variáveis contábeis das empresas de médio e baixo desempenho, quando à subamostra do 1º tercil eram agregadas, alternativamente, as subamostras do 2º e 3º tercis. A hipótese nula do teste é que os coeficientes de resposta sejam iguais entre os diferentes grupos.

A Tabela 3 apresenta os resultados dos testes dessas hipóteses, considerando que primeiramente foram testadas as hipóteses em que os coeficientes de respostas das variáveis das empresas de alto desempenho operacional quinquenal são iguais aos coeficientes das variáveis das empresas de médio desempenho. Posteriormente, testaram-se as hipóteses entre empresas de alto e baixo desempenhos.

Quando analisados os *betas* entre as empresas de alto e médio desempenhos, com base na evidência amostral e nível de significância de 5%, verifica-se que não existem razões para rejeitar a hipótese nula para as variáveis ΔRSPL e REC, o que indica que o mercado faz distinção, apenas, para a variação do retorno sobre investimento em termos de fluxo de caixa (ΔRIFC) e para a Estrutura de Capital (EC), enquanto que os demais coeficientes são considerados estatisticamente iguais.

No entanto, quando a comparação é realizada entre os *betas* das empresas de alto e baixo desempenhos, percebe-se que o mercado não faz distinção, apenas, entre a variável Estrutura de Capital, demonstrando que o mercado acionário brasileiro, ao precificar suas ações, diferencia as empresas de alto e baixo desempenho operacional de longo prazo, quando medido pelas variáveis contábeis deste estudo. Estes achados confirmam o estudo de Galdi e Lopes (2008), que encontrou evidências de um relacionamento entre o lucro e os preços das ações quando analisados no longo prazo.

Assim, é bastante provável que o investidor que tivesse escolhido uma estratégia de investir em ações de empresas com alto desempenho operacional de longo prazo, nesse estudo considerado em quinquênios móveis ao longo de 10 anos e medido por um mix ponderado envolvendo as variáveis contábeis RSPL, RICF, REC e EC, obteria retornos anormais positivos nos mesmos prazos quinquenais.

### *RESULTADO DOS TESTES DE ROBUSTEZ*

Para aferir a robustez dos resultados empíricos, foram realizados testes quanto à verificação de qual modelo de dados em painel é mais adequado – se efeitos fixos ou aleatórios, quanto à existência de raízes unitárias nas séries, quanto à distribuição das séries e à presença de autocorrelação e heterocedasticidade nos resíduos.

Tabela 3 - *Resultado do Teste de Wald.*
Table 3 – *Wald Test results.*

| Variáveis | Alto versus Médio | | Alto versus Baixo | |
|---|---|---|---|---|
| | Estatística F | p-valor | Estatística F | p-valor |
| ΔRSPL | 1,2934 | 0,1277 | 6,8398 | 0,0046 |
| ΔRIFC | 4,9306 | 0,0135 | 4,3640 | 0,0186 |
| REC | 1,0256 | 0,1559 | 5,6809 | 0,0086 |
| EC | 2,8572 | 0,0458 | 0,6971 | 0,2019 |

Foi realizado o Teste de *Hausman* para testar a hipótese de endogeneidade do termo aleatório e verificar qual o melhor modelo entre os efeitos fixos e aleatórios que apresenta resultados mais consistentes. Tendo em vista a dinâmica do presente estudo e os resultados do *Teste de Hausman* (P-valor = 0,0003), o modelo considerado mais consistente para os procedimentos é o modelo de efeitos fixos.

Os testes de raízes unitárias das séries entre as variáveis RSPL, RIFC, REC e EC foram baseados no Teste de *Fisher*, o qual assume um processo individual de raízes unitárias, e as probabilidades são computadas usando uma distribuição Qui-Quadrado. Os valores das estatísticas dos testes foram confrontados com a tabela dos valores de referência, confirmando-se, assim, que as séries REC e EC não apresentam raízes unitárias, porém, os testes sugerem não estacionariedade para as séries RSPL e RIFC, indicando que estas são integradas de 1ª ordem. A não estacionariedade de RSPL e RIFC pode estar relacionada à própria característica do lucro, que é suavizado em sua oscilação pela lógica do regime de competência e pelos *accruals* contábeis que podem ser gerenciados pelos gestores (Dechow *et al.*, 1998).

O Teste de *Jarque-Bera* tem o objetivo de testar se as séries são distribuídas normalmente. Sob a hipótese nula de uma distribuição normal, o teste revelou que as variáveis em estudo não possuem distribuição normal. No entanto, isso não invalida os resultados, visto que a amostra utilizada nesses modelos é suficientemente grande e pode-se recorrer ao Teorema do Limite Central. Conforme Brooks (2008), para amostras suficientemente grandes, violações das premissas de normalidade não provocam grandes consequências, utilizando-se o Teorema do Limite Central, as estatísticas do teste assintoticamente atingirão as distribuições adequadas, mesmo na ausência de normalidade.

Para testar a autocorrelação nos termos de perturbação, foi adotado o teste de *Durbin-Watson* (DW), o qual consiste em um teste de primeira ordem, isto é, testa apenas relação entre o termo do erro e seu valor imediatamente anterior. A hipótese nula deste teste é que os erros no tempo t e t-1 são independentes. Os testes demonstraram evidências de autocorrelação. Portanto, para obter resultados mais adequados, utilizou-se a técnica de *White-Period* (erro-padrão robusto), para a correção de autocorrelação serial e possíveis problemas de heterocedasticidade (*Eviews* 5, 2004). Essa é uma técnica adicional e que não altera os coeficientes *betas*, mas altera os valores dos coeficientes de erro-padrão e os *t* estatísticos das variáveis que são corrigidos diretamente nas matrizes da regressão.

A técnica de *White* pode ser usada quando se utiliza regressão com Mínimos Quadrados Ordinários. Assim, possíveis problemas são suprimidos ajustando o erro-padrão e o *t* estatístico, não violando os pressupostos da regressão. Erros-padrão robustos são frequentemente utilizados nos trabalhos com *cross section*, especialmente quando a amostra é grande (Wooldrigde, 2006).

**CONSIDERAÇÕES FINAIS**

Pesquisas sobre relevância da informação contábil para o valor da empresa, geralmente, avaliam o impacto do desempenho, medido por informações contábeis, no valor de mercado das empresas. Seguindo essa linha de pesquisa, este estudo propôs verificar se o mercado diferencia o desempenho operacional de longo prazo das empresas, medido por informações contábeis de diferentes naturezas, que sintetizam o desempenho quinquenal em empresas de alto, médio e baixo desempenho. A classificação das empresas nos três grupos de desempenho quinquenal foi realizada a partir de um índice síntese que atribui peso de 50% à variável retorno sobre o patrimônio líquido (RSPL) e distribui os outros 50% entre as variáveis retorno sobre o investimento gerado pelo fluxo de caixa das operações (RICF – 20%), variação quinquenal da receita (REC – 20%) e estrutura de capital (EC – 10%).

Os testes econométricos foram realizados pela técnica de regressão multivariada para dados em painel com efeitos fixos e variáveis *dummies*, em uma amostra composta por 142 empresas de diferentes setores listadas na Bolsa de Valores e de Mercadorias e Futuros de São Paulo (BM&FBOVESPA), no período de 1996 a 2009. Os resultados demonstram que o conjunto das variáveis contábeis utilizadas neste estudo para denotar o desempenho operacional das empresas é significantemente relevante para explicar o retorno das ações. Além disso, a redução consistente entre os grupos de alto, médio e baixo desempenho no coeficiente de resposta de cada variável contábil com esse retorno sugere evidências de que a escolha metodológica de construção do índice síntese de desempenho é robusta.

Os resultados sinalizam fortes evidências de que o mercado efetivamente diferencia as empresas de alto desempenho operacional de longo prazo. Como esperado, em termos gerais houve redução gradual no coeficiente de resposta das variáveis RSPL, RICF, REC e EC em sua associação com o retorno das ações quando a subamostra das empresas de alto desempenho se juntou, alternadamente, às subamostras de empresas de médio e baixo desempenho. Entretanto, a realização complementar do teste de *Wald* indicou que o mercado só diferencia, em termos estatísticos, as empresas de maior desempenho e pior desempenho.

Em um cenário em que o mercado é eficiente na forma semiforte, as informações geradas pela contabilidade deveriam estar refletidas nos preços das ações. Nessa situação, haveria uma forte relação entre a contabilidade e o mercado. Assim, como consequência dos resultados, observa-se que uma estratégia de investimento, *ex ante*, centrada nas empresas de maior desempenho de longo prazo, medidas pelas variáveis contábeis e metodologia definidas nesta pesquisa, pode levar a maiores retornos até que o próprio mercado se reequilibre nas condições de retorno normal. Sugere-se que este ponto seja explorado em futuras pesquisas.

Como limitação do estudo, cabe ressaltar que foram consideradas apenas empresas não financeiras que apresentassem ciclos operacionais de, no máximo, um ano. Com isso excluiu-se grande parte de ações do mercado acionário brasileiro. Para novas pesquisas, outra sugestão é replicar os métodos utilizados tomando como amostras empresas listadas na Bolsa de Valores de Nova Iorque e comparar com empresas negociadas na BM&FBOVESPA, para verificar o comportamento do retorno das ações frente ao desempenho operacional de longo prazo, em um mercado acionário desenvolvido e um emergente.

## REFERÊNCIAS

**MEG SARKIS SIMÃO ROSA**
Centro Universitário de Brasília
SEPN 707/907, Campus do Uniceub
Asa Norte
70790-075, Brasília, DF, Brasil

**PAULO ROBERTO BARBOSA LUSTOSA**
Universidade de Brasília
Faculdade de Economia, Administração e Contabilidade
Campus Universitário Darcy Ribeiro
70910-900, Brasília, DF, Brasil